\documentclass[prl,twocolumn,showpacs,tightenlines,superscriptaddress,
groupeaddress,preprintnumbers,amsmath,amssymb]{revtex4}
\usepackage{graphicx}

\newcommand{\beq}{\begin{equation}}
\newcommand{\eeq}{\end{equation}}
\newcommand{\bea}{\begin{eqnarray}}
\newcommand{\eea}{\end{eqnarray}}
\newcommand{\ba}{\begin{array}}
\newcommand{\ea}{\end{array}}
\newcommand{\bc}{\begin{center}}
\newcommand{\ec}{\end{center}}

\newcommand{\etal}{{\it et al.}}
\newcommand{\ibid}{{\it ibid. }}
\newcommand{\commentout}[1]{{}}
\newcommand{\half}{\hbox{$\frac{1}{2}$}}

\newcommand{\bk}{{\bf k}}

\newcommand{\eq}[1]{(\ref{#1})}
\newcommand{\bml}{\begin{subequations}}
\newcommand{\eml}{\end{subequations}}
\newcommand{\vol}[1]{{\bf #1}}
\newcommand{\comment}[1]{{}}

\begin{document}
\title{Cross-Molecular Coupling in Combined Photoassociation and Feshbach Resonances}

\author{Matt Mackie}
\affiliation {Department of Physics, Temple University, Philadelphia, PA 19122}
\affiliation{Department of Physics, University of Connecticut, Storrs, CT 06268}
\author{Matthew Fenty}
\affiliation {Department of Biology, Temple University, Philadelphia, PA 19122}
\affiliation {Department of Physics, Temple University, Philadelphia, PA 19122}
\author{Danielle Savage}
\affiliation {Department of Physics, Temple University, Philadelphia, PA 19122}
\author{Jake Kesselman}
\affiliation {Department of Physics, Temple University, Philadelphia, PA 19122}
\date{\today}

\begin{abstract}
We model combined photoassociation and Feshbach resonances in a Bose-Einstein condensate. When the magnetic field is far-off resonance, cross coupling between the two target molecules--enabled by the shared dissociation continuum--leads to an anomalous dispersive shift in the position of laser resonance, as well as unprecedented elimination and enhancement of resonant photoassociation via quantum interference. For off-resonant lasers, a dispersive shift and quantum interference appear similarly in resonant three-body Feshbach losses, except that the Feshbach node is tunable with intensity.
\end{abstract}


\maketitle

{\it Introduction.}--Photoassociation occurs when a pair of atoms absorb a laser
photon and thereby jump from the free-atom continuum to a bound molecular state~\cite{THO87}. At the turn of the last century, it was predicted that photoassociation could convert a condensate of atoms into a condensate of molecules~\cite{DRU98} which, in turn, raised the question of a rate limit on atom-molecule conversion in a condensate. The rate limit on photoassociative atom-molecule conversion arises either from two-body unitarity~\cite{BOH99}, or many-body rogue photodissociation to noncondensate atom pairs~\cite{KOS00,JAV02}. In the unitary limit, the DeBroglie wavelength sets the length scale $\ell=\Lambda_D$, whereas in the rogue limit the length scale is set by the interparticle distance $\ell=\rho^{-1/3}$, where $\rho$ is the system density. In either case, the fastest a molecular condensate can be created is $\sim m\ell^2/\hbar$.

Early condensate photoassociation experiments focused on bulk molecule formation~\cite{WYN00}, but next-generation experiments turned to the strongly interacting regime and the rate limit on atom-molecule formation~\cite{MCK02,PRO03}. Experiments with a Na condensate at NIST were thwarted by strong dipole forces~\cite{MCK02}, limiting the available laser intensity. Despite an intensity $\sim1$~kW/cm$^2$, the rate limit remained out of reach, whereas an intensity-dependent redshift of the photoassociation resonance was measured to be consistent with previous theory~\cite{FED96} and nondegenerate experiments~\cite{JON97}. The experiments at Rice focused on $^7$Li~\cite{PRO03}, and a laser intensity $\sim80$~W/cm$^2$ was sufficient to achieve a rate limit consistent with unitarity. However, the system was only borderline quantum degenerate, and the rogue limit could not be ruled out.

To probe the rate limit deeper, a Feshbach resonance was combined with photoassociation~\cite{JUN08}. Also known as magnetoassociation, a Feshbach resonance occurs when one atom from a colliding pair spin flips in the presence of a properly tuned magnetic field~\cite{STW76} and, similar to photoassociation, the pair jumps from the free-atom continuum to a bound molecular state. A magnetoassociation resonance enables a tunable interatomic scattering length~\cite{STW76,TOM98,INO98}, and can alleviate condensate instability problems~\cite{COR00}. Moreover, the Feshbach resonance is known to enhance~\cite{COU98,VUL99} or suppress~\cite{VUL99} photoassociation losses, and enhancement in particular could discern between the rate limits. In addition to an anomalous shift of laser resonance that is blue (red) for magnetic fields below (above) resonance, the latest experiments on Feshbach-assisted photoassociation of a condensate observe a rate constant that essentially vanishes ($\sim10^{-12}$~cm$^3$/s) at a particular below-resonance magnetic field and saturates on-resonance at an unprecedented value ($\sim10^{-7}$~cm$^3$/s)~\cite{JUN08}.

This Letter develops a simple analytical model of combined photoassociation and Feshbach resonances. Models presently are based on the idea of photoassociation with a Feshbach-tunable interatomic scattering length~\cite{COU98,VUL99,JUN08}, and apply only for off-resonant magnetic fields. Herein the basic quasi-continuum model is akin to two-color laser spectroscopy, valid in principle for simultaneous resonance. Moreover, while our analytical model ultimately encompasses models of photoassociation with a Feshbach-tunable scattering length~\cite{COU98,VUL99,JUN08}, and agrees reasonably with observation~\cite{JUN08}, we also foretell results for magnetoassociation with a laser-tunable scattering length. An experimental distinction between the unitary and rogue limits therefore remains elusive.

\begin{figure}[t]
\centering
\includegraphics[width=8.5cm]{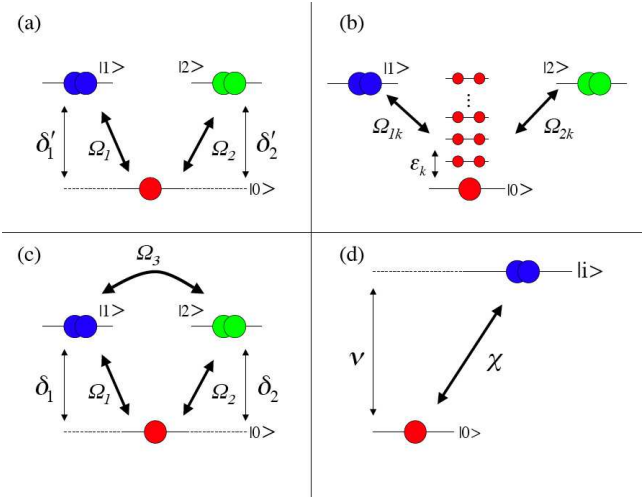}
\caption{(color online)~Few-level scheme for a condensate tuned nearby a combined photoassociation and Feshbach resonance. (a)~Basic three-level scheme, where a photoassociation and Feshbach resonance couple the atomic condensate $|0\rangle$ and molecular the condensates $|1\rangle$ and $|2\rangle$, respectively. (b)~A quasicontinuum accounts for dissociation to noncondensate levels. (c)~Eliminating the noncondensate levels leads to an effective $V$-system, where the virtual continuum couples the two molecular states, and where the detunings $\delta_i$ include the free-bound redshift. (d)~When the system is far from one resonance, magnetic or laser, the off-resonant molecular state can also be eliminated, leaving an effective two-level system, where the detuning $\nu$ includes an anomalous Stark-shift.}
\label{FEWL2}
\end{figure}

{\it Model.}--Consider $N$ atoms that have Bose condensed into, say, the zero momentum plane-wave ($\hbar\bk=0$)
state $|0\rangle$. Photoassociation and the Feshbach resonance then couple atoms in the state $|0\rangle$ to diatomic molecules of zero momentum in the states $|1\rangle$ and $|2\rangle$, respectively. As per Fig.~\ref{FEWL2}(a), this is the $V$-system familiar from few-level laser spectroscopy. Annihilation of an atom (molecule) of mass $m$ ($M=2m$) from the atomic ($i$th molecular) condensate is represented by the second-quantized operator $a_0\equiv a$ ($b_i$). This theory is the simplest, and molecules dissociate only back to the level $|0\rangle$. To be more complete, molecular dissociation to noncondensate levels should also be included [Fig.~\ref{FEWL2}(b)]. These levels must be considered because a condensate molecule need not dissociate back to the atomic condensate, but may just as well create a pair of atoms with equal-and-opposite momentum, since total momentum is conserved. So-called rogue
\cite{KOS00,JAV02}, or unwanted \cite{GOR01}, dissociation to noncondensate
modes therefore introduces the operators $a_{\pm\bk}$, as well as the kinetic energy
$\hbar\varepsilon_\bk=\hbar^2k^2/2m_r$ of an atom pair of reduced mass $m_r=m/2$.

To obtain mean-field equations, the Heisenberg equation for a given
operator is derived from the Hamiltonian (not shown), $i\hbar\dot{x}=[x,H]$, and all operators are subsequently declared $c$-numbers. In a minimalist model, $x$ represents either the atomic ($i$th molecular) operator $a_\bk$ ($b_i$), or the anomalous density operator $A_\bk=a_\bk a_{-\bk}$, where $A_\bk$ arises from rogue
dissociation to noncondensate atom pairs of equal-and-opposite momentum. Converting summations over $\bk$ to integrals over frequency $\varepsilon$ introduces the characteristic frequency $\omega_\rho=\hbar\rho^{2/3}/2m_r$. The corresponding mean-field model is ($i=1,2$)
\bml
\bea
i\dot{a} &=& -\Omega_1 a^* b_i -\Omega_2 a^* b_2, 
\\
i\dot{b}_i &=&  \tilde\delta_i' b_i
  -\half\Omega_i a^2-\half\xi_i\!\int\!d\varepsilon\sqrt{\varepsilon}\,f_i(\varepsilon)A(\varepsilon),
\label{BDOT}
\\
i\dot{A}(\varepsilon) &=&\varepsilon A(\varepsilon)
    -\Omega_1\,f_1(\varepsilon)b_1-\Omega_2\,f_2(\varepsilon)b_2.
\label{ADOT}
\eea
\label{BOSE_EQM}
\eml
The amplitudes are of unit order, the $i$th atom-molecule coupling is $\Omega_i\propto\sqrt{\rho}$, the rogue dissociation coupling is $\xi_i=\Omega_i/(4\pi^2\omega_\rho^{3/2})$, the frequency dependence of the $i$th atom-molecule coupling is $f_i(\varepsilon)$, and the tunable binding energy of the $i$th molecular state
is $\hbar\delta_i'$. Lastly, spontaneous decay of the photoassociation molecule has been
included as $\tilde\delta_1'=\delta_1'-i\Gamma_s/2$, and spontaneous decay of the
Feshbach molecule~\cite{KOE05} has been neglected.

It remains to model the continuum by specifying the shapes $f_i(\varepsilon)$. If a condensed-matter-type universality is to be the driving paradigm then the only lengthscale in the strongly-interacting problem is the interatomic distance, and a single theta-function, $f_i=\Theta(\varepsilon-\beta_i)$ with $\beta_i=\omega_\rho$, could be employed. However, universality has already failed in magnetoassociation of $^6$Li~\cite{JAV05}, where the proper length scale was the size of the Feshbach molecule, and we therefore allow for different length scales with the Lorentzian $f_i^2=1/(1+4\varepsilon^2/\beta_i^2)$.

{\it Effective Few-Level Systems.}--Here our focus will be on the situation where only one field, laser or magnetic, is resonant, and the other is off resonant. Numerically, Eqs.~\eq{BOSE_EQM} are stiff for large detunings ($\delta_i'\gg\Omega_i$), and an analytical solution is therefore enabled by eliminating the rogue amplitude using $\dot{A}\approx0$; Eq.~\eq{ADOT} yields $A(\varepsilon)=(\Omega_1f_1b_1+\Omega_2f_2b_2)/\varepsilon$, which is put into Eq.~\eq{BDOT}, and dissociation is then re-introduced phenomenologically. This simple approximation is related to the Fermi Golden Rule and, like the Golden Rule, arguably reveals the essential physics. The model~\eq{BOSE_EQM} becomes
\bml
\bea
i\dot{a} &=& -\Omega_1 a^* b_1 -\Omega_2 a^* b_2, 
\\
i\dot{b}_1 &=&  \tilde\delta_1 b_1
  -\half\Omega_1 a^2-\half\Omega_3b_2,
\\
i\dot{b}_2 &=&  \tilde\delta_2 b_2
  -\half\Omega_2 a^2-\half\Omega_3b_1,
\label{B2DOT}
\eea
\label{VCAS_EQM}
\eml
where $\tilde\delta_1=\delta_1-i(\Gamma_s+\Gamma_1)/2$, 
$\tilde\delta_2=\delta_2-i\Gamma_2/2$, and where
$\Omega_3=\Omega_1\Omega_2/(4\pi^2\omega_\rho^{3/2})\int d\varepsilon
  f_1(\varepsilon)f_2(\varepsilon)/\sqrt{\varepsilon}
    \approx(\eta\Omega_1\Omega_2/4\pi\omega_\rho)\sqrt{\beta_1/\omega_\rho}\,$; note, $\eta=0.8346$ is leftover from a hypergeometric function in the limit of point-like Feshbach molecules ($\beta_2\gg\beta_1$). The detunings have the usual~\cite{FED96} redshift $\delta_i=\delta_i'-\Sigma_i$, where
 $\Sigma_i=\Omega_i^2/(8\pi^2\omega_\rho^{3/2})\int d\varepsilon
f_i^2(\varepsilon)/\sqrt{\varepsilon}=\Omega_i^2/(16\pi\omega_\rho)\sqrt{\beta_i/\omega_\rho}$.  The dissociation rates are 
$\Gamma_i=\Omega_i^2/(8\pi\omega_\rho)\sqrt{\epsilon_i/\omega_\rho}$,
with $\hbar\epsilon_i$ the energy of the atom pair~\cite{YUR03}.
Incidentally, $\Sigma_i$ and $\Gamma_i$ are the real and imaginary parts of the Lamb shift from laser spectroscopy. Last, the shared continuum acts like a virtual state that couples the photoassociation and Feshbach molecules with Rabi frequency $\Omega_3$ [Fig.~\ref{FEWL2}(c)].
 
Now, starting with an off-resonant magnetic field 
($\delta_2\gg\Omega_2,\Gamma_2$), we take $\dot{b}_2/\delta_2\approx0$ in 
Eq.~\eq{B2DOT}; hence, $b_2=(\Omega_2a^2+\Omega_3b_1)/2\delta_2$ is substituted into Eqs.~\eq{VCAS_EQM}, yielding the two-mode system [Fig.~\ref{FEWL2}(d) with $|i\rangle=|1\rangle$]: $i\dot{a}=-\chi a^*b_1$, $i\dot{b}_1=(\nu-i\Gamma/2) b_1-\half\chi a^2$,
where $\chi=\Omega_1+2{\cal L}\delta_2\Omega_2/\Omega_3$, $\nu=\delta_1-{\cal L}\delta_2$, and $\Gamma=\Gamma_s+\Gamma_1+{\cal L}\Gamma_2$, where ${\cal L}=\Omega_3^2/4|\tilde\delta_2|^2$. Note that, although the usual singularities are absent for $\delta_2=0$, the model is still only valid far-off resonance ($\delta_2\gg\Omega_2,\Gamma_2$). Beyond the usual Feshbach mean-field shift, $|a|^2\Omega_2^2\delta_2/2|\tilde\delta_2|^2$, which has been neglected compared to $\Gamma_s$~\cite{JUN08}, we find a ``cross-molecular" shift of the laser resonance, $\Sigma_3=-{\cal L}\delta_2$, that is blue (red) below (above) the Feshbach resonance. Just as surprising is the fact that effective photoassociation ceases, i.e., $\chi=0$, for $\delta_2\approx-\Omega_2\Omega_3/2\Omega_1$. Borrowing again from laser spectroscopy, this node arises from destructive interference between direct photoassociation and photoassociation via the Feshbach molecular state. Similarly, constructive interference occurs above resonance, but observation is complicated by condensate instability due to a negative resonant scattering length~\cite{JUN08}.

On the other hand, we may also consider an off-resonant laser 
($\delta_1\gg\Omega_1,\Gamma_s,\Gamma_1$), which leads to essentially the same two-mode system [Fig.~\ref{FEWL2}(d) with $|i\rangle=|2\rangle$]: $i\dot{a}=-\chi a^*b_2$, $i\dot{b}_2=(\nu-i\Gamma/2) b_2-\half\chi a^2$, except that 
$\chi=\Omega_2+2{\cal L}\delta_1\Omega_1/\Omega_3$, $\nu=\delta_2-{\cal L}\delta_1$, and 
$\Gamma=\Gamma_s+\Gamma_1+{\cal L}\Gamma_2$, where 
${\cal L}=\Omega_3^2/4|\tilde\delta_1|^2$. In addition to a dispersive shift ${\cal L}\delta_1$, we find that $\chi=0$ and atom losses due to magnetoassociation, i.e., {\em three-body losses near a Feshbach resonance}, vanish for laser detuning $\delta_1\approx-\Omega_1\Omega_3/2\Omega_2$. The Feshbach node of course arises from destructive interference between direct magnetoassociation and magnetoassociation via the photoassociation state. Moreover, whereas the strength of a given Feshbach resonance--and thus the detuning position of the above photoassociation node--is fixed, the laser intensity can be varied. For arbitrary red detuning $-\delta_1\agt 10\,\Gamma_s$ and $\Omega_{1(3)}=\bar\Omega_{1(3)}\sqrt{I}$, the intensity position of the Feshbach node is $I_n=-2\delta_1\Omega_2/(\bar\Omega_1\bar\Omega_3)$.

{\it Comparison to Observation.}--To compare with the latest observations~\cite{JUN08}, we first need the rate equation for losses from the atomic condensate~\cite{KOS00}: 
$\dot{P}_0=-\rho KP_0^2$, where $P_0=|a|^2$ and 
$\rho K=\half\chi^2\Gamma/(\nu^2+\Gamma^2/4)$ defines the rate constant $K$. Next,  the photoassociation parameters $\Omega_1$, $\epsilon_1$, and $\beta_1$ are approximated by comparing to previous experiments~\cite{PRO03}, where the natural molecular linewidth is $\Gamma_s=12\times2\pi$~MHz, using the definitions $\Omega_1=\bar\Omega_1\sqrt{I}$, $\epsilon_1=p_1^2/2m_r$ and $\beta_1=\hbar/(2m_rL_\beta^2)$. In the low-intensity limit, we find  $\bar\Omega_1\sqrt{I_0}=290\times2\pi$~kHz, with the saturation intensity $I_0=28$~W/cm$^2$ defined by $\Gamma_1/\Gamma_s=I/I_0$. From the high-intensity limit, the characteristic momentum is $p_1= 2.21\,\hbar/\lambdabar$, roughly twice the photon recoil momentum for light of wavelength $2\pi\lambdabar=671$~nm. From the lightshift, $L_\beta=116a_0$ is comparable to the classical size of the photoassociation molecule~\cite{PRO03,JUN08} ($a_0$ is the Bohr radius). The main parameters for the Feshbach coupling are the product of the zero-field scattering length and the resonance width~\cite{JUN08}, $|a_{\rm bk}|\Delta_B=1.6$~nm$\cdot$G, and the difference in magnetic moments between the Feshbach molecule and the free-atom pair~\cite{JUN08}, $\Delta_\mu=2\mu_0$ (the Bohr magneton is $\mu_0$), so that $\Omega_2=\sqrt{8\pi\rho|a_{\rm bk}|\Delta_B\Delta_\mu/m}=127\times2\pi$~kHz. The cross-molecular coupling is then $\Omega_3=\bar\Omega_3\sqrt{I}$, where $\bar\Omega_3\sqrt{I_0}=138\times2\pi$~MHz. The Feshbach detuning is related to the magnetic field by 
$\hbar\delta_2=(B-B_0)\Delta_\mu$, where $B_0=736$~G~\cite{JUN08} locates resonance. The sole unfixed parameter is the kinetic energy of the magnetodissociated pair, $\hbar\epsilon_2=p_2^2/2m_r$. For a zero-temperature homogeneous system of point-like Feshbach molecules, the reasonable ansatz for the length scale is $p_2= \hbar\rho^{1/3}$.

\begin{figure}[b]
\centering
\includegraphics[width=8.5cm]{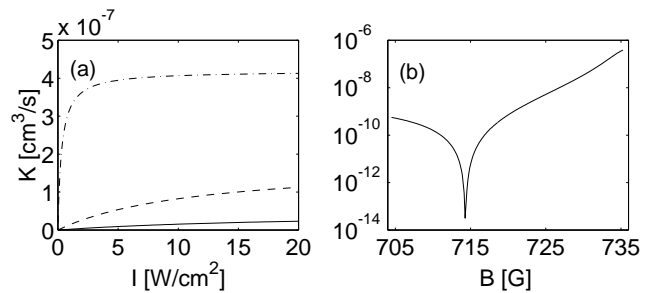}
\caption{Laser-resonant rate constant vs.~(a) intensity and (b)~magnetic field for photoassociation of a $^7$Li condensate tuned nearby a Feshbach resonance. In panel (a), the rate constant saturates at a limit and intensity that depends on magnetic field, where the solid (dashed, dash-dotted) line is for $B=728$~G (732~G, 735.5~G). In panel (b), the rate constant for $I=10$~W/cm$^2$ approaches the result for photoassociation alone at low fields, essentially vanishes for $B=714$~G, and rises to unprecedented values near resonance. Note that the model is on the edge of validity at $B=735.5$~G.}
\label{K_CROSS}
\end{figure}

Results are shown in Fig.~\ref{K_CROSS} for condensate density $\rho=10^{12}\,{\rm cm}^{-3}$ and a laser tuned to lightshifted resonance. The net lightshift per unit intensity is $\Sigma'=(\Sigma_1+\Sigma_3)/I=(\Gamma_s/2I_0)\sqrt{\beta_1/\gamma_1}+\bar{\cal L}\delta_2$ with $\bar{\cal L}=\bar\Omega_3^2/4|\tilde\delta_2|^2$; for $B=732$~G, the net lightshift is blue $\Sigma'=-13\times2\pi$~MHz/(W/cm$^2$), in reasonable agreement with observation~\cite{JUN08}. Far from the Feshbach resonance, laser-resonant losses approach those for photoassociation alone [panel~(a), solid line]; but, as the Feshbach resonance is approached, the saturation intensity decreases and the rate limit increases [panel~(a), dashed and dash-dotted lines]. From the definition $\Gamma=\Gamma_s+\Gamma_1+{\cal L}\Gamma_2$, saturation sets in when $\Gamma_1+{\cal L}\Gamma_2\sim\Gamma_s$, which translates into a saturation intensity $I_1/I_0=1/(1+\bar{\cal L}I_0\Gamma_2/\Gamma_s)$; for $B=732$~G, we find $I_1=12$~W/cm$^2$. As for the rate limit, it is roughly the rate for converting atoms into Feshbach molecules, $R_0\approx2\Omega_2^2/\Gamma_2=16\pi\omega_\rho$. This estimate is best near resonance: for $B=732$~G and 735.5~G, exact results from the definition of $K$ are $R_0/16\pi\omega_\rho=0.35$ and 0.84, respectively. The $B=732$~G results are in reasonable agreement with Ref.~\cite{JUN08}. Lastly, losses cease for $\delta_2\approx-\Omega_2\Omega_3/2\Omega_1$, or $B=714$~G [Fig.~\ref{K_CROSS}(b)], which is intensity independent but depends on the classical size of the photoassociated molecule through $\Omega_3\propto L_\beta$, again in reasonable agreement with experiments~\cite{JUN08}.

For a magnetodissociation momentum set by the interparticle distance, theory compares reasonably with observation. However, since the off-resonant size of the Feshbach molecule is roughly the resonant scattering length, and since $\rho^{-1/3}\sim a_{\rm res}(732\:{\rm G})$~\cite{JUN08}, the correct length scale is somewhat ambiguous. We expect that modelling near-Feshbach resonance experiments will require the classical size of the Feshbach molecule. Finally, in that the saturation limit is set by the rate for magnetoassociation alone, $16\pi\hbar\rho^{2/3}/m$, these results agree with the rogue model~\cite{JAV02} up to a dimensionless constant. However, the observed unitary limit~\cite{JUN08} coincides with the rogue model herein at $B=732$~G, and a definitive distinction remains elusive. Also, future work is needed to determine if the rate of Feshbach-assisted photoassociation of a condensate saturates or maximizes.

Before closing, we make the connection to existing models~\cite{VUL99,JUN08} of photoassociation near a Feshbach resonance. For an off-resonant magnetic field (recall, $\delta_2\gg\Omega_2,\Gamma_2$), the Feshbach-resonant interatomic scattering length is defined $4\pi\hbar\rho a_{\rm res}/m=-\Omega_2^2/2\delta_2$, so that the effective coupling becomes $\chi=\Omega_1(1-\eta a_{\rm res}/L_\beta)$. The photoassociation node occurs when the Feshbach-resonant scattering length equals the classical size of the photoassociated molecule, $a_{\rm res}\approx L_\beta$. Not incidentally, what appears to be the correct definition of $a_{\rm res}$ leads, ultimately, to the $16\pi$ in the rate limit. Similarly, the net lightshift can be written $\Sigma=\Sigma_1(1-2\eta^2a_{\rm res}/L_\beta)$, which crosses zero, i.e., blue to-and-from red, at about the same detuning position as the photoassociation node. Of course, the atom-atom scattering length can also be tuned with off-resonant photoassociation~\cite{THE04,FED96,KOS00}, and analogous results apply to the Feshbach nodes.

{\it Conclusion.}--We have reported a general model for a Bose-Einstein condensate near a combined photoassociation and Feshbach resonance.  When the magnetic field is far-off resonance, cross molecular coupling between the two target molecules leads to an anomalous dispersive shift in the position of laser resonance. Moreover, even though the magnetic field is far from resonance, and the probablility of forming Feshbach molecules is miniscule, this cross-coupling can eliminate or enhance resonant photoassociation via quantum interference. Unfortunately, a definitive experimental distinction between the unitary and rogue limits for photoassociation of a condensate remains elusive. Nevertheless, for a far-detuned laser, a similar shift and interference arises in resonant Feshbach losses, except that the interference is tunable with laser intensity. These results are typical of two-color laser spectroscopy, despite the presence of only a single laser.

{\it Acknowledgements.}--We gratefully acknowledge helpful conversations with Randy Hulet, Juha Javanainen, Steve Gensemer, Mark Junker, and Dan Dries, as well as support from the NSF (MM, PHY-0354599), NSF-AMP~(DS), and the Temple University Office of the Vice Provost for Undergraduate Affairs~(JK).

\end{document}